\documentclass[preprint, aps, pra, showpacs, floatfix, superscriptaddress]{revtex4-1}

\usepackage{longtable}
\usepackage{graphicx}
\usepackage{amsmath,amsfonts,amssymb,verbatim}
\usepackage[usenames]{color}    
\usepackage{colortbl}
\usepackage{calrsfs}
\usepackage{ulem} 
\usepackage{hyperref} 
\usepackage{subfigure}
\usepackage{indentfirst}
\usepackage{tikz} 
\usepackage{pgfplots}
\graphicspath{{graph/}}
\definecolor{faded}{gray}{0.45}
\newcommand{\bp}{{\bf p}}
\newcommand{\bk}{{\bf k}}
\newcommand{\br}{{\bf r}}

\newcommand{\ba}{{\bf a}}
\newcommand{\bY}{{\bf Y}}
\newcommand{\balpha}{\boldsymbol{\alpha}}

\newcommand{\bepsilon}{\boldsymbol{\epsilon}}
\begin{document}
\thispagestyle{empty}
\title{Electronic structure effects in the electron bremsstrahlung from heavy ions}
\author{M.~E.~Groshev}
\affiliation{
Department of Physics, St. Petersburg State University,
Universitetskaya Naberezhnaya 7/9, 199034 St. Petersburg, Russia
}
\author{V.~A.~Zaytsev}
\affiliation{
Department of Physics, St. Petersburg State University,
Universitetskaya Naberezhnaya 7/9, 199034 St. Petersburg, Russia
}
\author{V.~A.~Yerokhin}
\affiliation{
Center for Advanced Studies, Peter the Great St.~Petersburg Polytechnic University,
St. Petersburg 195251, Russia
}
\author{P.-M.~Hillenbrand}
\affiliation{
GSI Helmholtzzentrum f\"ur Schwerionenforschung,
64291 Darmstadt, Germany
}
\affiliation{Institut f\"ur Kernphysik, Goethe-Universit\"at, 
60438 Frankfurt, Germany}

\author{Yu.~A.~Litvinov}
\affiliation{
GSI Helmholtzzentrum f\"ur Schwerionenforschung,
64291 Darmstadt, Germany
}
\author{V.~M.~Shabaev}
\affiliation{
Department of Physics, St. Petersburg State University,
Universitetskaya Naberezhnaya 7/9, 199034 St. Petersburg, Russia
}
%
%
\begin{abstract}
A fully relativistic approach is presented for the calculation of the bremsstrahlung emitted by
an electron scattered off an ionic target.
The ionic target is described as a combination of an effective Coulomb potential and a finite-range
potential induced by the electronic cloud of the ion.
The approach allows us to investigate the influence of the electronic structure of the target
on the properties of the emitted radiation.
We calculate the double differential cross-section and Stokes parameters of the bremsstrahlung
of an electron scattered off uranium ions in different charge states, ranging from bare to neutral uranium.
%
%
Results on the high-energy  endpoint of the electron bremsstrahlung from Li-like uranium ions $\mathrm{U}^{89+}$ are compared to the recent experimental data.
%
%
For this process, it is found that taking into account the electronic structure of the target results in modification of the cross-section on the level of 14\%, which can, in principle, be seen in present-day experiments.
\end{abstract}
%
%
\maketitle
%
%
\section{INTRODUCTION}
%
%
Electron bremsstrahlung is a process of the photon emission induced by the electron deceleration in the field of the ionic or atomic target. 
Investigations of this process are both of fundamental and of practical interest since they can provide information about the electronic structure and polarization properties of the target (see, e.g., reviews~\cite{Mangiarotti_RPC141_312:2017, Jakubassa_arXiv_2103.06034}).
%
%
%
\\ \indent
Experimentally, bremsstrahlung is most studied for neutral atomic targets. 
The state-of-the-art theoretical calculations rely on the partial-wave representation of the Dirac electron wave function in the field of a finite-range central potential. 
This approach provides results in a remarkable agreement with experiment~\cite{ 
TsengPratt_PRA3_100:1971,
TsengPratt_PRA7_1502:1973,
Yerokhin_PRA82_062702:2010,
Tashenov_PRA87_022707:2013,
Muller_PRA90_032707:2014,
Poskus_CPC04_030:2018, 
Poskus_ADN129_101277:2019}. 
%
%
For incident electrons with energies above a few hundred keV, the dominant contribution to the bremsstrahlung comes from the scattering off the nuclear field, whereas the contribution of the electronic cloud is small. 
In this case, the field of the atomic target is often replaced with the pure Coulomb potential~\cite{
Jakubassa_PRA82_042714:2010, 
Jakubassa_PRA98_062715:2018,
Jakubassa_PRA100_032703:2019, 
Jakubassa_JPG47_075102:2020}. 
At higher collision energies starting from several MeV and large momentum transfers (corresponding to large photon emission angles), the extended size of the nucleus becomes significant and needs to be taken into account.
The corresponding theoretical calculations are performed with the Dirac wave functions in the combined Coulomb potential and a finite-range potential caused by the nuclear charge distribution
\cite{
Jakubassa_PRA93_052716:2016,
Jakubassa_arXiv_2103.06034}.
\\ \indent
Recently it became possible to study the bremsstrahlung in collisions of electrons with highly-charged ions~\cite{
Nofal_PRL99_163201:2007, 
Hillenbrand_PRA90_022707:2014,
Hillenbrand_PRA101_022708:2020}. 
Since it is difficult to form a target made of highly-charged ions, the actual study is performed in the inverse kinematics. 
Namely, a beam of highly charged ions collides with a cloud of light neutral atoms. 
During the collision, the quasifree electron of the target is ionized and captured into the continuum of the heavy-ion projectile, with simultaneous emission of a photon. 
This process is called the radiative electron capture to the continuum (RECC). 
In the projectile rest frame, the RECC process is approximately equivalent to the bremsstrahlung in which the emitted radiation takes away the entire energy of the incoming electron, the so-called high-energy endpoint of the bremsstrahlung spectrum.
\\ \indent
So far, theoretical studies of the RECC process (see calculations in Refs.~\cite{
Nofal_PRL99_163201:2007, 
Hillenbrand_PRA90_022707:2014, 
Hillenbrand_PRA101_022708:2020})
were carried out under the assumption that the field of the heavy ion is represented by the pure Coulomb potential. 
In such approach, the influence of the ion's electronic structure on the properties of the emitted radiation is neglected. 
However, for collision energies around tens of keV used in actual experiments~\cite{
Nofal_PRL99_163201:2007, 
Hillenbrand_PRA90_022707:2014,
Hillenbrand_PRA101_022708:2020},
the electronic structure of the projectile may yield a sizeable effect on the properties of the emitted radiation. 
Here we work out a formalism for calculating the bremsstrahlung from an electron scattered off a combination of the Coulomb potential and a finite-range potential induced by the electron cloud of the
ion.
\\ \indent
Our formalism can be viewed as an extension of one developed by Yerokhin and Surzhykov~\cite{Yerokhin_PRA82_062702:2010}, where the cases of the pure Coulomb and a finite-range potential were implemented. 
The particular form of the potential is important since it defines the asymptotic behaviour of the Dirac wave function at large radial distances. 
In order to perform radial integrations with highly oscillating functions, we use the method of the complex-plane rotation of the integration contour, which relies on the analytical properties of the Dirac wave functions in the asymptotic region.
The Dirac solutions for a finite-range potential at large distances are represented by spherical Bessel functions, which is the simplest case for the complex-plane rotation. 
For the pure Coulomb potential, the Dirac wave functions are
described by the regular at origin solution of the Dirac-Coulomb equation, i.e., the Whittaker function of the first kind $M_{\alpha\beta}$. 
For the sum of the Coulomb and a finite-range potential, the Dirac wave functions at large distances are represented by a linear superposition of the regular and irregular Dirac-Coulomb wave functions. 
This significantly complicates the numerical implementation of the complex-plane rotation method of computation of radial integrals.
\\ \indent
In the present work, we apply the developed method for calculating the bremsstrahlung emitted by electrons scattered off uranium ions. 
By comparing the results obtained for different charge states of the ion, we study the effects of the electronic structure on the angular distribution and polarization of the emitted photons. 
We also calculate the triple differential cross-section of RECC for Li-like uranium studied experimentally in Ref.~\cite{Hillenbrand_PRA101_022708:2020}.
The theoretical results are found to be in good agreement with the experimental data. 
It is found that the electronic-structure effects modify the cross-section of this process by 14\%. 
%
This is comparable with the experimental uncertainty of Ref.~\cite{Hillenbrand_PRA101_022708:2020} but can be detected in dedicated future experiments.
\\ \indent
The paper is organized as follows. In Sec.~\ref{basics} we recall basic relations for the electron bremsstrahlung. 
The double differential cross-section and the Stokes parameters are discussed in Secs.~\ref{ddcs} and~\ref{Stokes}, respectively. 
Section~\ref{experiment} is devoted to the comparison of the obtained results with the experimental data from Ref.~\cite{Hillenbrand_PRA101_022708:2020}. 
Finally, a summary is given in Sec.~\ref{conclusion}.
\\ \indent
Relativistic units $(m_e = \hbar = c = 1)$ and the Heaviside charge units $(e^2 = 4\pi\alpha)$ are utilized throughout the paper.
%
%
\section{BASIC FORMALISM}
\label{basics}
%
%
A fully differential cross-section of the electron bremsstrahlung in the field of the ionic or atomic target is given by
\begin{equation}
\frac{d\sigma_{\bp_f\mu_f,\, \bk\lambda;\; \bp_i\mu_i}}{d\omega d\Omega_k d\Omega_{p_f}}
=
(2\pi)^4 \omega^2 \frac{p_f \varepsilon_f\varepsilon_i}{p_i}
|\tau_{\bp_f\mu_f,\, \bk\lambda;\; \bp_i\mu_i}|^2,
\label{eq:tdcs}
\end{equation}
where $\varepsilon$, $\bp$ and $\mu$ are the energy, asymptotic momentum, and polarization of the electron in the initial ($i$) and final ($f$) states, $p = |\bp|$, $(\omega, \bk)$ is the four-momentum of the emitted photon with the polarization $\lambda$.
%
The amplitude is expressed by
\begin{equation}
\tau_{\bp_f\mu_f,\, \bk\lambda;\;\bp_i\mu_i}  =
\int d^3\br
\Psi^{(-)\dagger}_{\bp_f \mu_f}(\br)
\hat R_{\bk \lambda}^\dagger(\br)
\Psi^{(+)}_{\bp_i \mu_i}(\br).
\label{eq:amplitude}
\end{equation}
Here the photon emisison operator is defined as follows
\begin{equation}
\hat{R}^\dagger_{\bk\lambda}(\br)
 =
-\sqrt{\frac{\alpha}{(2\pi)^2\omega}}
\balpha\cdot\bepsilon_{\lambda}^*
e^{-i\bk\cdot \br},
\label{eq:photon_operator}
\end{equation}
where $\balpha$ stands for the vector of Dirac matrices and the Coulomb gauge fixes the circular polarization vector $\bepsilon_{\lambda}$.
The wave functions of the incoming $\Psi_{\bp\mu}^{(+)}$ and outgoing $\Psi_{\bp\mu}^{(-)}$ electrons are given by~\cite{Rose_1961, Pratt_RMP45_273:1973, Eichler_1995}
\begin{equation}
\Psi_{\bp\mu}^{(\pm)}(\br)
 =
\frac{1}{\sqrt{4\pi p\varepsilon}}
\sum\limits_{\kappa m_j}
i^l
e^{\pm i \delta_{\kappa}}
\sqrt{2l+1}
C^{j\mu}_{l0\,1/2\mu}
D^j_{m_j\mu}(\varphi_{p}, \theta_{p}, 0)
\Psi_{\varepsilon \kappa m_j}(\br).
\label{eq:wf_in_out}
\end{equation}
Here
$\kappa = (-1)^{l+j+1/2}\left(j + 1/2\right)$ is the Dirac quantum number determined by the angular momentum $j$ and the parity $l$,
$m_j$ is the angular momentum projection,
$\delta_{\kappa}$ is the phase shift induced by the scattering potential,
$C^{JM}_{j_1 m_1\,j_2 m_2}$ is the Clebsch-Gordan coefficient,
$D^J_{MM'}$ is the Wigner matrix~\cite{Rose_1957, Varshalovich},
and the azimuthal $\varphi_{p}$ and polar $\theta_{p}$ angles define the direction of the momentum~$\bp$.
The partial waves
\begin{equation}
\Psi_{\varepsilon\kappa m_j}(\br)
=
\frac{1}{r}
\begin{pmatrix}
G_{\varepsilon\kappa}(r)\Omega_{\kappa m_j}(\hat\br)
\\
iF_{\varepsilon\kappa}(r)\Omega_{-\kappa m_j}(\hat\br)
\end{pmatrix}
\label{eq:partial_wave}
\end{equation}
are the continuum Dirac eigenstates in the scattering central potential $V(r)$, where $G_{\varepsilon\kappa}$ and $F_{\varepsilon\kappa}$ are the large and small radial components, $\Omega_{\kappa m_j}$ is the spherical spinor~\cite{Varshalovich}, and $\hat\br = \br/|\br|$.
In the present study, $V(r)$ is an effective (Coulomb and screening) potential created by the nucleus and target electrons
\begin{equation}
V(r)
=
V_{\rm nucl}(r)
+
V_{\rm scr}(r)
\xrightarrow[r\rightarrow +\infty]{}
-\frac{\alpha Z_{\rm as}}{r},
\label{eq:central_pot}
\end{equation}
where $Z_{\rm as} = Z - N_{e}$ with $N_{e}$ being the number of target electrons.
We note that the constructed wave functions of the incoming and outgoing electrons~\eqref{eq:wf_in_out} effectively take into account the electron-target interaction in a nonperturbative manner.
\\ \indent
The multipole expansion of the photon field~\cite{Rose_1961} is
\begin{equation}
\bepsilon_\lambda^* e^{-i\bk\br}
=
\sqrt{2\pi}
\sum\limits_{LM_L}
i^{-L}
\sqrt{2L+1}
D^{L*}_{M_L\lambda}(\varphi_{k},\theta_{k},0)
\sum\limits_{p=0,1}
(-i\lambda)^p \ba_{LM_L}^{(p)*}(\br),
\label{eq:multipole_expansion}
\end{equation}
where $\ba_{LM_L}^{(p)}$ are the magnetic ($p=0$) and electric ($p=1$) vectors
\begin{equation}
\begin{aligned}
\ba_{LM_L}^{(0)}(\br)
&=
j_L(\omega r)\bY_{LLM_L}(\hat\br),
\\
\ba_{LM_L}^{(1)}(\br)
&=
\sqrt{\frac{L+1}{2L+1}}
j_{L-1}(\omega r)\bY_{LL-1M_L}(\hat\br)
-
\sqrt{\frac{L}{2L+1}}
j_{L+1}(\omega r)\bY_{LL+1M_L}(\hat\br)
\end{aligned}
\label{eq:magn_elec_vec}
\end{equation}
with $j_L$ standing for the spherical Bessel function of the first kind~\cite{Abramovitz}
and $\bY_{JLM}$ are the vector spherical harmonics~\cite{Varshalovich}.
\\ \indent
Substituting Eqs.~\eqref{eq:photon_operator} and \eqref{eq:wf_in_out} into Eq.~\eqref{eq:amplitude} and utilizing Eqs.~\eqref{eq:partial_wave} and ~\eqref{eq:multipole_expansion}, we obtain the bremsstrahlung amplitude expressed as an infinite triple sum over the partial waves and photon multipoles:
\begin{equation}
\begin{aligned}
\tau_{\bp_f\mu_f,\, \bk\lambda;\; \bp_i\mu_i}
&=
\frac{1}{4\pi}
\sqrt{\frac{\alpha}{2\pi\omega}}
\frac{1}{\sqrt{p_i\varepsilon_ip_f\varepsilon_f}}
\sum\limits_{\kappa_im_i}
i^{l_i}
e^{i\delta_{\kappa_i}}
\sqrt{\frac{2l_i+1}{2j_i+1}}
C^{j_i\mu_i}_{l_i0\;1/2\mu_i}
D^{j_i}_{m_i\mu_i}(\varphi_{p_i},\theta_{p_i},0)
\\
&\times
\sum\limits_{\kappa_fm_f}
i^{-l_f}
e^{i\delta_{\kappa_f}}
\sqrt{2l_f+1}
C^{j_f\mu_f}_{l_f0\;1/2\mu_f}
D^{j_f*}_{m_f\mu_f}(\varphi_{p_f},\theta_{p_f},0)
\\
&\times
\sum\limits_{LM_L}
(-i)^L
\sqrt{2L+1}
C^{j_im_i}_{j_f m_f\;LM_L}
D^{L*}_{M_L\lambda}(\varphi_{k},\theta_{k}, 0)
\sum\limits_{p=0,1}
(-i\lambda)^p
{\langle \varepsilon_i \kappa_i||\balpha\cdot \ba^{(p)*}_{L}||\varepsilon_f \kappa_f\rangle}.
\end{aligned}
\label{eq:full_amplitude}
\end{equation}
Here ${\langle \varepsilon_i \kappa_i||\balpha\cdot \ba^{(p)*}_{L}||\varepsilon_f \kappa_f\rangle}$ are the reduced matrix elements whose explicit form is presented in Appendix~\ref{appendix_rme}.
Numerical evaluation of the reduced matrix elements is a difficult task.
For the pure Coulomb potential, the method of calculation of these elements was worked out in Ref.~\cite{Yerokhin_PRA82_062702:2010}.
Here we expand this algorithm to the case of the central potential being a superposition of the Coulomb and short-range potentials.
The developed approach is discussed in details in Appendix~\ref{appendix_rf}.
%
%
%
%
\section{NUMERICAL CALCULATIONS}
\label{results}
%
%
%
%
Here we concentrate on the evaluation of the bremsstrahlung from electrons scattered off uranium ($Z=92$) targets in various charge states.
Interest in this system is formed by a series of performed~\cite{Nofal_PRL99_163201:2007, Hillenbrand_PRA90_022707:2014, Hillenbrand_PRA101_022708:2020} and planned experiments.
We restrict ourselves to the case when the incident electron is spin-unpolarized, and only the emitted photon is detected.
For such a scenario, all characteristics of the process are defined by the double differential cross-section (DDCS)
\begin{equation}
d\sigma_{\lambda}
\equiv
\frac{d\sigma_{\lambda;\; \bp_i}}{d\omega d\Omega_k}
=
\frac{1}{2}\sum_{\mu_i\mu_f}
\int d\Omega_{p_f}
\frac{d\sigma_{\bp_f\mu_f,\, \bk\lambda;\; \bp_i\mu_i}}{d\omega d\Omega_k d\Omega_{p_f}}.
\label{eq:ddcs}
\end{equation}
%
%
%
\subsection{Double differential cross-section}
\label{ddcs}
\begin{figure}[h!]
\begin{center}
\includegraphics[width=\linewidth]{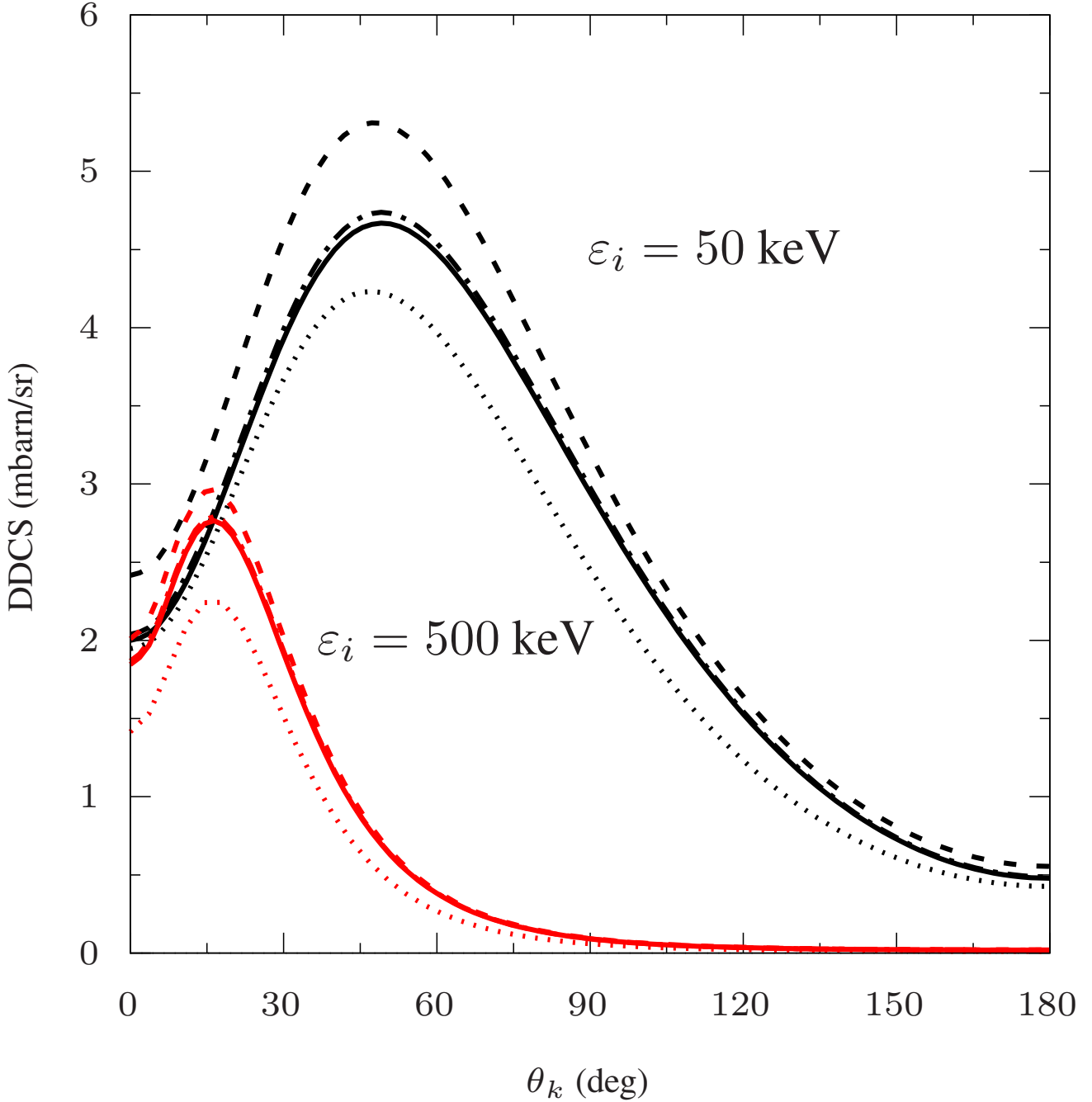}
\caption{
Normalized double differential cross-section~\eqref{eq:normalaized_ddcs} as a function of the photon emission angle $\theta_k$.
The incident electron with the kinetic energy $\varepsilon_i = 50$ keV (black) and $500$ keV (red) is scattered off the Ne-like uranium ${\rm U}^{82+}$, which is described by the Coulomb potential with $Z_{\rm as} = 82$ (dotted), core-Hartree (solid), and Perdew-Zunger (dash-dotted) potentials.
The results for the bare uranium ${\rm U}^{92+}$ (dashed) are presented as well.
The photon energy $\omega = 0.99\varepsilon_i$.}
\label{fig:scr_vs_coul}
\end{center}
\end{figure}

\begin{figure}[h!]
\begin{center}
\includegraphics[width=0.5\linewidth]{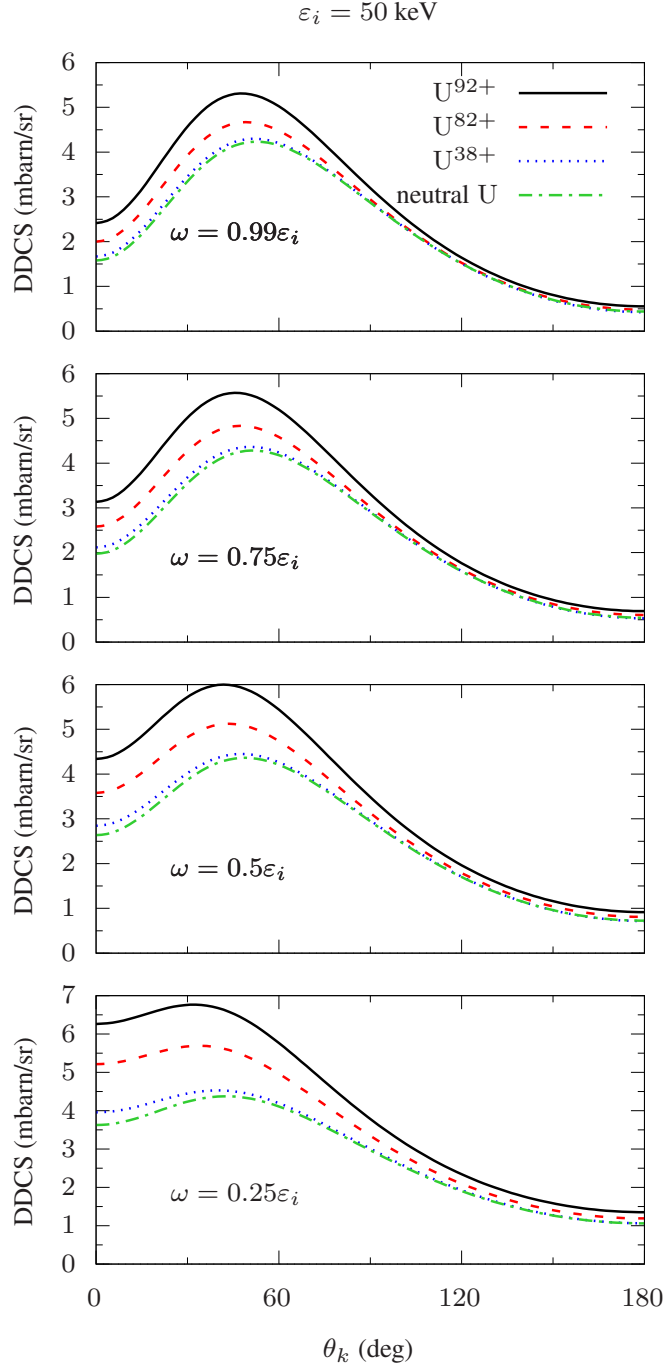}
\caption{
Normalized double differential cross-section~\eqref{eq:normalaized_ddcs} as a function of the photon emission angle $\theta_k$.
The results are represented for the bare $\mathrm{U^{92+}}$ (black solid), Ne-like $\mathrm{U^{82+}}$ (red dashed), Xe-like $\mathrm{U^{38+}}$ (blue dotted), and neutral (green dash-dotted) uranium targets.
The incident electron kinetic energy is $\varepsilon_i = 50$ keV.
The photon energies $\omega = 0.99\varepsilon_i$, $0.75\varepsilon_i$, $0.5\varepsilon_i$, and $0.25\varepsilon_i$ are presented in the first, second, third, and fourth rows, respectively.}
\label{fig:ddcs_50}
\end{center}
\end{figure}

\begin{figure}[h!]
\begin{center}
\includegraphics[width=0.5\linewidth]{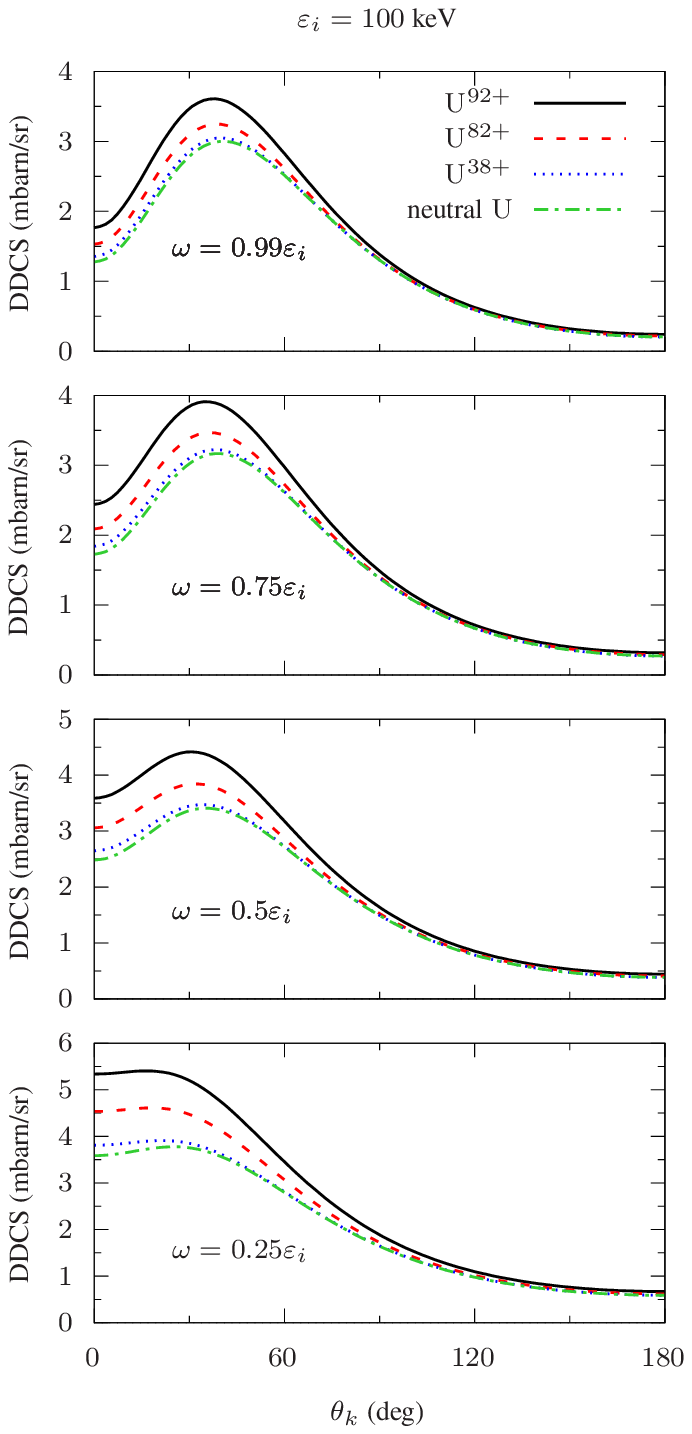}
\caption{
Normalized double differential cross-section~\eqref{eq:normalaized_ddcs} as a function of the photon emission angle $\theta_k$.
The results are represented for the bare $\mathrm{U^{92+}}$ (black solid), Ne-like $\mathrm{U^{82+}}$ (red dashed), Xe-like $\mathrm{U^{38+}}$ (blue dotted), and neutral (green dash-dotted) uranium targets.
The incident electron kinetic energy is $\varepsilon_i = 100$ keV.
The photon energies $\omega = 0.99\varepsilon_i$, $0.75\varepsilon_i$, $0.5\varepsilon_i$, and $0.25\varepsilon_i$ are presented in the first, second, third, and fourth rows, respectively.}
\label{fig:ddcs_100}
\end{center}
\end{figure}

\begin{figure}[h!]
\begin{center}
\includegraphics[width=0.5\linewidth]{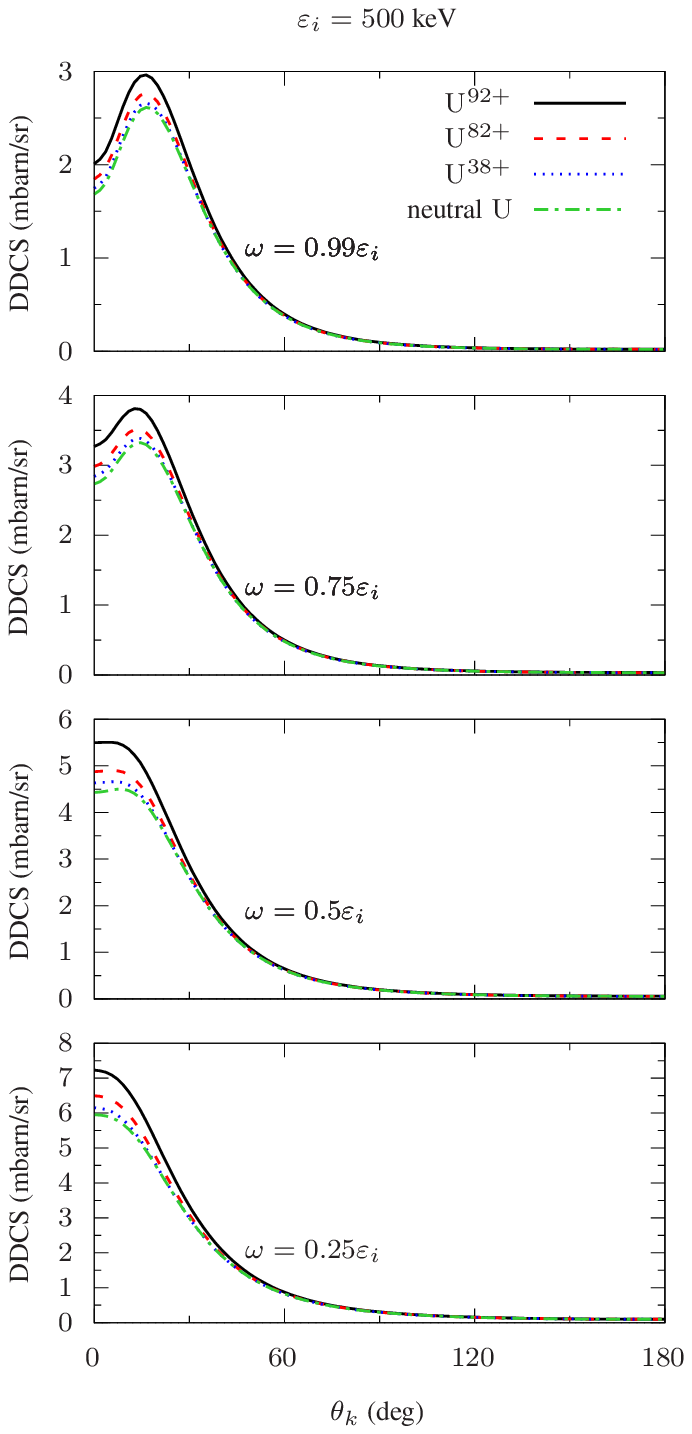}
\caption{
Normalized double differential cross-section~\eqref{eq:normalaized_ddcs} as a function of the photon emission angle $\theta_k$.
The results are represented for the bare $\mathrm{U^{92+}}$ (black solid), Ne-like $\mathrm{U^{82+}}$ (red dashed), Xe-like $\mathrm{U^{38+}}$ (blue dotted), and neutral (green dash-dotted) uranium targets.
The incident electron kinetic energy is $\varepsilon_i = 500$ keV.
The photon energies $\omega = 0.99\varepsilon_i$, $0.75\varepsilon_i$, $0.5\varepsilon_i$, and $0.25\varepsilon_i$ are presented in the first, second, third, and fourth rows, respectively.}
\label{fig:ddcs_500}
\end{center}
\end{figure}

%
%
We start by studying the dependence of the DDCS on the choice of the effective screening potential.
Here, we consider Perdew-Zunger (PZ)~\cite{PerdewZunger} and core-Hartree (CH) potentials, with the latter one defined as
\begin{equation}
V^{\rm (CH)}_{\rm scr}(r)
=
\alpha
\int\limits_0^{+\infty} dr'
\frac{\rho_{e}(r')}{r_>},
\end{equation}
where $r_> = \max(r, r')$, the charge density is given by
\begin{equation}
\rho_{e}(r)
=
\sum\limits_{i=1}^{N_{e}}
\Bigl[
G^2_{n_i\kappa_i}(r) + F^2_{n_i\kappa_i}(r)
\Bigr],
\label{eq:CH_density}
\end{equation}
and normalized to the number of electrons in the target
\begin{equation}
\int\limits_{0}^{+\infty}dr
\rho_{e}(r)
=
N_{e}.
\end{equation}
In Eq.~\eqref{eq:CH_density}, the summation is carried out over occupied states.
\\ \indent
The normalized DDCS
\begin{equation}
d\sigma
=
\frac{\omega}{Z^2}
\sum\limits_{\lambda=\pm 1}
d\sigma_\lambda
\label{eq:normalaized_ddcs}
\end{equation}
for the bremsstrahlung from $50$ and $500$ keV electrons in the field of the Ne-like uranium ${\rm U}^{82+}$, being described by different scattering potentials, is presented in Fig.~\ref{fig:scr_vs_coul}.
The energy of the emitted photon $\omega = 0.99\varepsilon_i$.
In such a case, when the photon carries away almost the whole energy of the incoming electron, the influence of the target electronic structure on the bremsstrahlung is the most pronounced.
And, as a result, the strong dependence of the DDCS on the choice of the screening potential is expected.
From Fig.~\ref{fig:scr_vs_coul}, it is seen that the difference between the DDCS, calculated with the CH and PZ scattering potentials, is negligible.
In what follows, therefore, we perform the calculations only with the core-Hartree potential.
Though the form of the screening potential almost doesn't affect the results, taking into account of the short-range potential leads to prominent changes in the DDCS, which correspond to a rather large a difference between the dotted and solid lines in the figure.
This difference amounts to about 10\% and 20\% for low and high kinetic energies of the incident electron.
%
%
\\
\indent
We now turn to the investigation of dependence of the bremsstrahlung DDCS~\eqref{eq:normalaized_ddcs} on the number of target electrons $N_e$.
In Figures~\ref{fig:ddcs_50}, \ref{fig:ddcs_100}, and \ref{fig:ddcs_500}, we present the results for the bremsstrahlung of $50$ keV, $100$ keV, and $500$ keV electrons, respectively, in the fields of the bare, Ne-like, Xe-like, and neutral uranium targets.
%
The scattering potential of the neutral atom is described by a sum of three Yukawa terms~\cite{Salvat_1987, Salvat_1991}
\begin{equation}
V_{\rm neut}(r)
=
-\frac{\alpha Z}{r}
\sum\limits_{i=1}^3
A_ie^{-\alpha_ir}
\label{eq:yukawa}
\end{equation}
with $A_i$ and $\alpha_i$ standing for the potential amplitude and scaling constant, respectively.
We note that our results for the neutral atom are in good agreement with ones from Ref.~\cite{Yerokhin_PRA82_062702:2010}, where completely different algorithms have been utilized.
%
%
%
%
From Fig.~\ref{fig:ddcs_50}, it is seen that for low energies of the incident electron, the dependence on the number of target electrons becomes the most prominent.
In contrast, at high energies, the differences in the DDCS for different charge states of the target become smaller.
This can be explained as follows.
The bremsstrahlung from high-energy electrons comes mainly from the region close to the nucleus, while the contribution from the target electrons is of minor importance.
For incident electrons with low energies, the region of the dominant contribution is shifted to distances that are closer to the spectator electrons and, as a result, the dependence on the number of the target electrons increases.
It should also be mentioned that the emmited photons with the energies $\omega = 0.99\varepsilon_i$ are mostly emitted in the course of bremsstrahlung from the nuclear region.
In this case, as can be seen from Figs.~\ref{fig:ddcs_50}, \ref{fig:ddcs_100}, and~\ref{fig:ddcs_500}, the curves corresponding to the different charge states of the target lay closer to each other when compared to other photon energies.
%
%
\subsection{Stokes parameters}
\label{Stokes}
%
%
To evaluate the influence of the target electronic structure on the bremsstrahlung polarization properties, we consider the Stokes parameters defined by
\begin{equation}
P_1
=
\frac{d\sigma_{0^{\circ}} - d\sigma_{90^{\circ}}}
{d\sigma_{0^{\circ}} + d\sigma_{90^{\circ}}},
\quad
P_2
=
\frac{d\sigma_{45^{\circ}} - d\sigma_{135^{\circ}}}
{d\sigma_{45^{\circ}} + d\sigma_{135^{\circ}}},
\quad
P_3
=
\frac{d\sigma_{+1} - d\sigma_{-1}}
{d\sigma_{+1} + d\sigma_{-1}}.
\label{eq:Stokes}
\end{equation}
Here $d\sigma_\chi$ stands for the DDCS~\eqref{eq:ddcs} of the linear polarized bremsstrahlung with the polarization angle $\chi$,
while $d\sigma_{+1}$ and $d\sigma_{-1}$ are the cross sections for the emission of right and left circularly polarized photons, respectively.
To obtain the expression for $d\sigma_\chi$, one needs to replace $\bepsilon_\lambda$ in~Eq.\eqref{eq:photon_operator} with the vector of the linear polarization
\begin{equation}
\bepsilon_\chi
=
\frac{1}{\sqrt{2}}
\sum\limits_{\lambda = \pm 1}
e^{i\lambda\chi}
\bepsilon_\lambda.
\end{equation}
Since the incident electron is spin-unpolarized, $P_2$ and $P_3$ are identically equal to zero~\cite{TsengPratt_PRA7_1502:1973}.
Therefore, only the degree of linear polarization $P_1$ is relevant.
\begin{figure}[h!]
\begin{center}
\includegraphics[width=1.\linewidth]{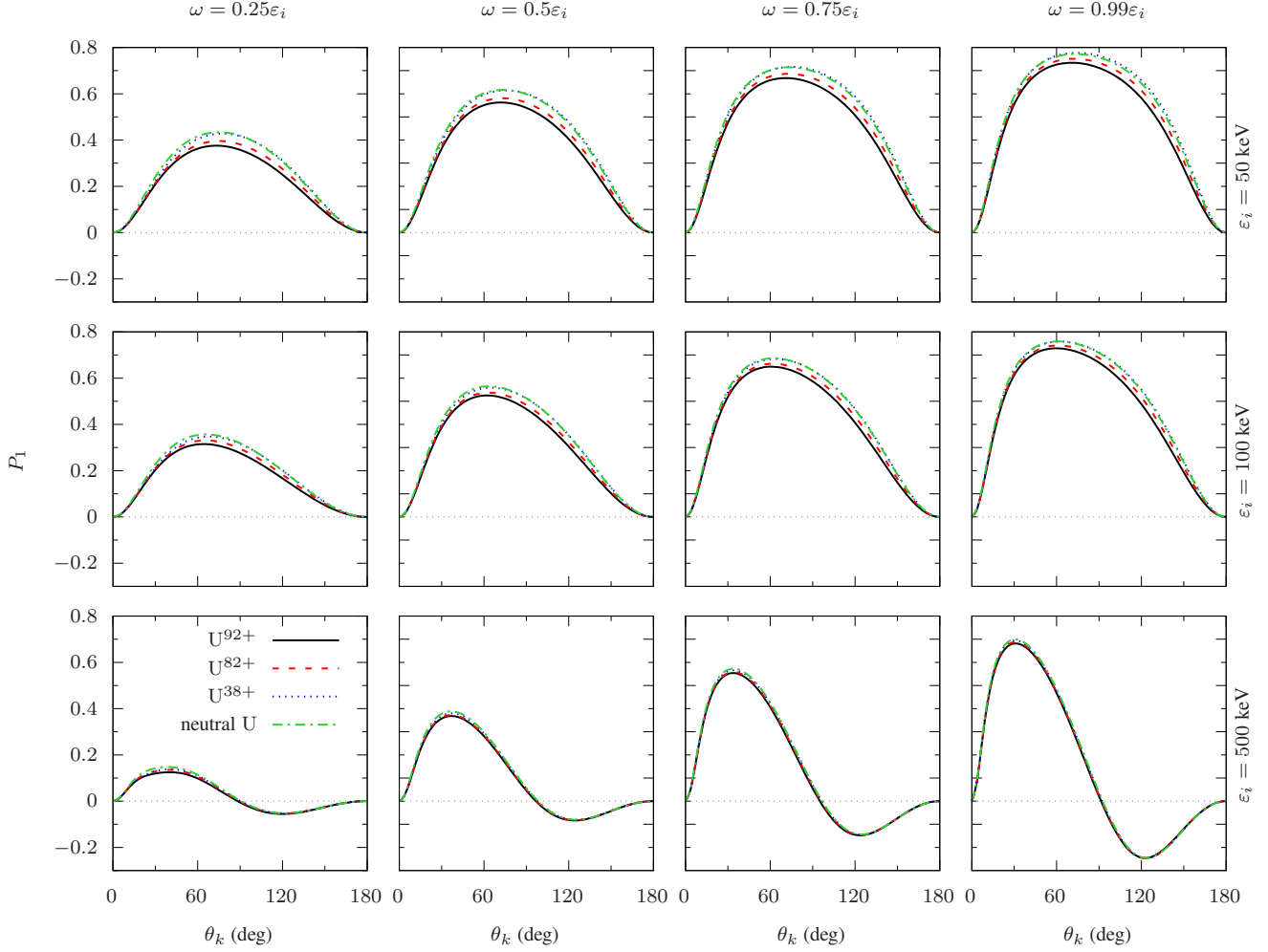}
\caption{
The degree of linear polarization $P_1$ as a function of the photon emission angle $\theta_k$.
The results are represented for the bare $\mathrm{U^{92+}}$ (black solid), Ne-like $\mathrm{U^{82+}}$ (red dashed), Xe-like $\mathrm{U^{38+}}$ (blue dotted), and neutral (green dash-dotted) uranium targets.
The first, second, and third rows correspond to the incident electron kinetic energies $\varepsilon_i = 50$ keV, $100$ keV, and $500$ keV, respectively.
The photon energies $\omega = 0.25\varepsilon_i$, $0.5\varepsilon_i$, $0.75\varepsilon_i$, and $0.99\varepsilon_i$ are presented in the first, second, third, and fourth columns, respectively.}
\label{fig:p1}
\end{center}
\end{figure}
In Figure~\ref{fig:p1}, the Stokes parameter $P_1$ is shown as a function of the photon emission angle $\theta_k$ for the same set of the parameters as in Figs.~\ref{fig:ddcs_50},~\ref{fig:ddcs_100}, and~\ref{fig:ddcs_500}.
From this figure, it is seen that the dependence of $P_1$ on the number of target electrons is barely recognizable.
%
%
\subsection{Radiative electron capture to the continuum}
\label{experiment}
\begin{figure}[h!]
\begin{center}
\includegraphics[scale=1.5]{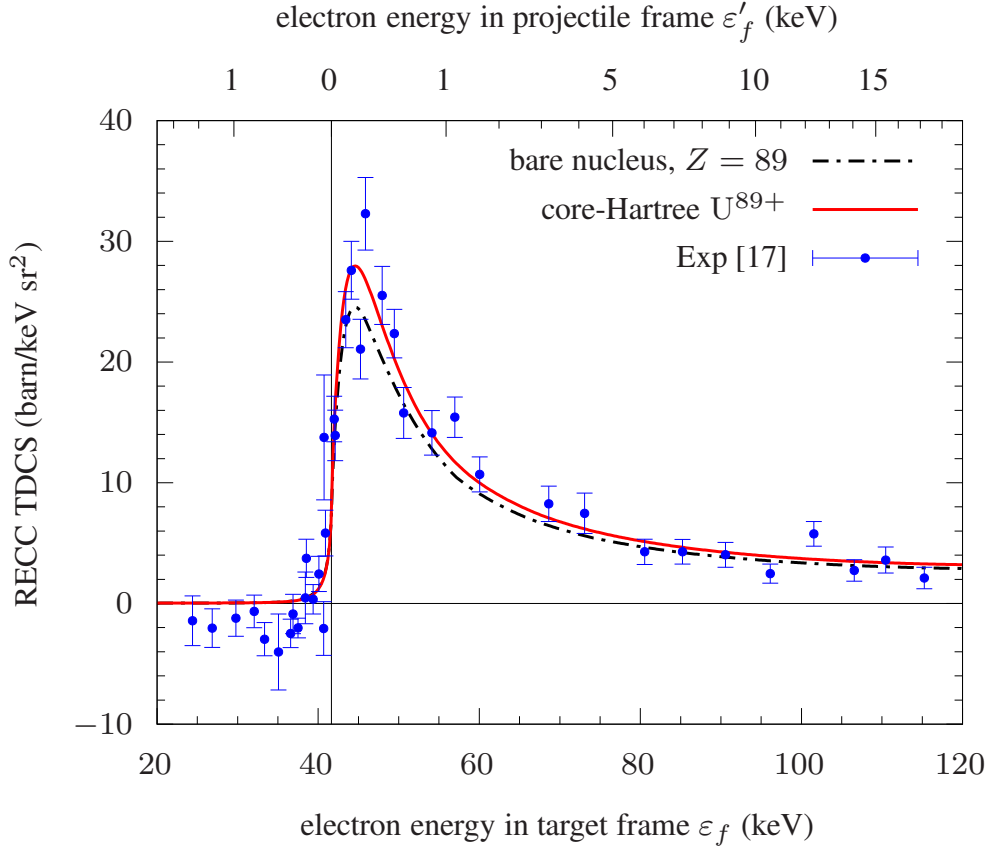}
\caption{
The RECC triple differential cross-section as a function of the measured electron energy.
The results are shown for ${\rm U}^{89+}$ described by the core-Hartree potential (red solid) and pure Coulomb potential with $Z = 89$ (black dash-dotted).
Experimental data from Ref.~\cite{Hillenbrand_PRA101_022708:2020} are represented by points with error bars signifying the statistical uncertainty.}
\label{fig:recc}
\end{center}
\end{figure}

%
%
In collisions of highly-charged heavy ions with light atoms or molecules, an ion can capture an electron from the target discrete spectrum.
If the electron is captured to the projectile continuum and one photon is emitted, the process is called radiative electron capture to the continuum (RECC).
A theoretical description of RECC was first presented by Jakubassa-Amundsen in a series of papers~\cite{Jakubassa_JPB36_1971:2003, Jakubassa_RPC75_1319:2006, Jakubassa_JPB40_2719:2007, Jakubassa_EPJD41_267:2007}.
Measurements of the RECC allow one to study electron bremsstrahlung at the high-energy endpoint.
At this endpoint, where the emitted photon carries away the highest possible energy, the bremsstrahlung emission exhibits the most strong dependence on the electronic structure of the projectile.
\\ \indent
With this in mind, we turn to the theoretical description of the RECC channel in the collision of Li-like uranium ions with a supersonic gas-jet target of molecular nitrogen:
\begin{equation}
\nonumber
\mathrm{U}^{89+}(1s^22s) + \mathrm{N}_2
\rightarrow
\mathrm{U}^{89+}(1s^22s) + \bigl[\mathrm{N}^{+}_2\bigr]^* + e^{-}(\varepsilon_f, \theta_{p_f}) + \gamma_{\rm ph}(\omega, \theta_k).
\end{equation}
This process was experimentally studied just recently in Ref.~\cite{Hillenbrand_PRA101_022708:2020}.
In the projectile reference frame, the RECC process is related to the high-energy endpoint of the bremsstrahlung from the target quasifree electrons scattered off the heavy projectile.
Utilizing this relation and approximating the field of the Li-like uranium by the pure Coulomb potential with the charge $Z = 89$, the theoretical description has been performed.
Such a description completely neglects the electronic structure of the projectile.
However, for considered energies of the incident quasifree electrons $\varepsilon'_i = 41.64$ keV (here and throughout, primes refer to the variables defined in the projectile reference frame), the electronic structure can provide a prominent contribution to the cross-section.
Here we evaluate this contribution with the usage of the developed algorithm.
\\ \indent
To compare the theoretical results with the measured $0^\circ$ electron spectra of the RECC one needs to average the bremsstrahlung TDCS over laboratory-frame polar $\theta_{p_f}$ and azimuthal $\varphi_{p_f}$ angles as follows~\cite{Hillenbrand_PRA90_022707:2014, Hillenbrand_PRA101_022708:2020}
%
\begin{equation}
\frac{d\sigma^{(\mathrm{RECC})}_{\bp_i}}{d\omega d\Omega_k d\Omega_{p_f}}\Bigg\vert_{\theta_{p_f} = 0^\circ}
=
\frac{Z_t}{\gamma^2(1-\beta\cos\theta_k)^2}
\int\limits_{0}^{\theta_{\rm max}}d\theta_{p_f}
\frac{\sin \theta'_{p_f}}{1 - \cos \theta_{\rm max}}
\int\limits_{0}^{2\pi}
\frac{d\varphi_{p_f}}{2\pi}
\sum\limits_{\lambda}
\frac{d\sigma^{(\mathrm{brem})}_{\lambda; \bp_i}}{d\omega' d\Omega'_k d\Omega'_{p_f}}.
\label{eq:tdcs_recc}
\end{equation}
Here $Z_t = 7$ is the nuclear charge of the target nitrogen atom, $\beta = 0.3808$ and $\gamma = 1.081$ are the velocity and Lorentz factors of the projectile, respectively, and the prefactor originates from the relation between the solid angles in the projectile $d\Omega'_k$ and laboratory $d\Omega_k$ frames (see, e.g., Ref.~\cite{Eichler_PR439_1:2007})
\begin{equation}
\frac{d\Omega'_k}{d\Omega_k} 
= 
\frac{1}{\gamma^2(1-\beta\cos\theta_k)^2}.
\end{equation}
The bremsstrahlung TDCS is averaged over the initial and summed over final polarizations and averaging over $0^\circ \leqslant \theta_{p_f} \leqslant \theta_{\rm max}$ and $0^\circ \leqslant \varphi_{p_f} < 360^\circ$ covered by the spectrometer is hold.
%
%
%
The emission angles of the scattered electron $\theta'_{p_f}$ and emitted photon $\theta'_{k}$ in the projectile frame are connected to the observation angles $\theta_{p_f}$ and $\theta_{k}$ in the target frame as follows
\begin{equation}
\theta'_{p_f} = \pi - \arctan\left[
\frac{\sin\theta_{p_f}}{\gamma(\cos\theta_{p_f}-\beta\varepsilon_f/p_f)}
\right],
\end{equation}
\begin{equation}
\theta'_{k} = \pi - \arctan\left[
\frac{\sin\theta_{k}}{\gamma(\cos\theta_{k}-\beta)}
\right].
\end{equation}
%
The relation of the projectile-frame energies $\varepsilon'_f$ and $\omega'$ to the corresponding observables in the target frame is given by the Lorentz transformation
\begin{equation}
\varepsilon'_f = \varepsilon'_i + \gamma\varepsilon_f-\gamma \beta p_f \cos\theta_{p_f},
\end{equation}
\begin{equation}
\omega' = \gamma\omega(1-\beta\cos\theta_k).
\end{equation}
We fix the photon emission angle $\theta_k = 90^\circ$ ($\theta'_k = 67.6^\circ$ in the projectile reference frame)
and the maximal polar angle $\theta_{\rm max} = 3.0^\circ$ to match the conditions in Ref.~\cite{Hillenbrand_PRA101_022708:2020}.
\\
\indent
In Figure~\ref{fig:recc}, the TDCS of RECC is presented as a function of the captured electron energy in the laboratory (target) frame.
The experimental TDCS measured on a relative scale were normalized to the theoretical results employing the core-Hartree potential.
From this figure, it is seen that the results obtained for the core-Hartree potential, which effectively accounts for the interaction of the quasifree electron with the Li-like uranium ion, differs from the ones obtained for the pure Coulomb potential.
Taking into account the projectile electronic structure leads to quite prominent changes in TDCS.
As the photon energy $\omega'$ approaches the initial kinetic energy of the electron $\varepsilon'_i$, the difference between two theoretical curves increases and reaches 14\% at the high-energy endpoint.
At lower photon energies, the deviation is about 10\%.
%
%
The uncertainty of the experimental data from [17] was dominated by the statistical uncertainty, which could be reduced straightforward in a future measurement.
Therefore, such a deviation can, in principle, be detected in experiment. 
This justifies the importance of the calculations accounting for the electron-target interaction in a non-perturbative regime.
%
%
\section{CONCLUSION}
\label{conclusion}
%
%
In this paper, we have presented the fully relativistic description of the electron bremsstrahlung in the field of ionic targets.
The electron-target interaction was treated non-perturbatively by using solutions of the Dirac equation for the combination of the Coulomb and the screening potentials for the incoming and outgoing electrons.
In the case of the pure Coulomb and a finite-range potential, our calculations were shown to agree with those reported in Ref.~\cite{Yerokhin_PRA82_062702:2010}.
\\
\indent
The developed approach was applied to the evaluation of the double differential cross-sections and Stokes parameters of the bremsstrahlung from the bare, Ne-like, Xe-like, and neutral uranium.
We found that for $50$ keV incident electrons taking into account of the target electronic structure alters the double differential cross-section by $10$-$20\%$ as compared to the case of the pure Coulomb potential.
In contrast, the linear polarization of the bremsstrahlung was shown to be insensitive to the electronic structure (i.e., the charge state) of the target.
\\
\indent
Furthermore, we have applied the developed method to the calculation of the radiative electron capture to the continuum in the collision of the Li-like uranium ions with molecular nitrogen, which was experimentally studied recently in Ref.~\cite{Hillenbrand_PRA101_022708:2020}.
It was demonstrated that accounting for the electronic structure of $\mathrm{U}^{89+}$ projectile results in significant changes of the triple differential cross-section.
The changes as high as $14\%$ were found in comparison with the results obtained for the pure Coulomb potential with the effective charge $Z = 89$ (used in Ref.~\cite{Hillenbrand_PRA101_022708:2020}). 
Effects of this magnitude can be detected in modern
experiments at heavy-ion Storage rings.
We conclude that the electronic structure effects should be taken into account in the theoretical description of the electron bremsstrahlung from the highly-charged ions in the energy region of a few tens of keV.
%
%
\section{ACKNOWLEDGMENTS}
%
%
The work was supported by the German–Russian Interdisciplinary Science Center (G-RISC Sur-Place stipend No.~F-2021b-25).
V.M.S. acknowledges the support by RFBR and ROSATOM according to the 
research project No.~20-21-00098.
%
%
\appendix
\section{REDUCED MATRIX ELEMENTS}
\label{appendix_rme}
%
%
Results for the reduced matrix elements in Eq.~\eqref{eq:full_amplitude} are
\begin{equation}
\begin{aligned}
(-i)
\langle
\varepsilon_i \kappa_i || \balpha\cdot \ba_{L}^{(0)*} || \varepsilon_f \kappa_f
\rangle
&=
P_{LL}(i,f),
\\
(-i)
\langle
\varepsilon_i \kappa_i ||  \balpha\cdot \ba_{L}^{(1)*}  || \varepsilon_f \kappa_f
\rangle
&=
\sqrt{\frac{L+1}{2L+1}}P_{LL-1}(i,f)
-
\sqrt{\frac{L}{2L+1}}P_{LL+1}(i,f).
\end{aligned}
\end{equation}
Here the radial integrals are given by
\begin{equation}
P_{JL}(i,f)
=
\int\limits_0^{+\infty}dr
j_L(\omega r)
\Big[
S_{JL}(\kappa_i, -\kappa_f)
G_{\varepsilon_i\kappa_i}(r)
F_{\varepsilon_f\kappa_f}(r)
-
S_{JL}(-\kappa_i, \kappa_f)
G_{\varepsilon_f\kappa_f}(r)
F_{\varepsilon_i\kappa_i}(r)
\Big],
\label{eq:radial_integral}
\end{equation}
where the angular coefficients $S_{JL}(\kappa_i, \kappa_f)$ are defined as in Ref.~\cite{Yerokhin_PRA82_062702:2010}
\begin{equation}
S_{JL}(\kappa_i; \kappa_f)
=
(-1)^{l_f}
\sqrt{\frac{3}{2\pi}}
\Pi_{j_i j_f l_i l_f J}
C^{L0}_{l_i0\;l_f0}
\begin{Bmatrix}
j_i & l_i & 1/2
\\
j_f & l_f & 1/2
\\
J & L & 1
\end{Bmatrix},
\label{eq:sjl}
\end{equation}
with $\Pi_{ab\ldots c} = \sqrt{(2a+1)(2b+1)\ldots(2c+1)}$ and $\{\ldots\}$ stands for $9j$-symbol~\cite{Varshalovich}.
%
%
\section{RADIAL FUNCTIONS}
\label{appendix_rf}
%
%
\indent
Outside the finite-range potential, the radial parts of the electron continuum wave functions are expressed through the solutions of the Dirac equation in the pure Coulomb potential with the effective asymptotic charge $Z_{\rm as}$ as follows
\begin{equation}
\left(
\begin{aligned}
G_{\varepsilon\kappa}
\\
F_{\varepsilon\kappa}
\end{aligned}
\right)
=
\cos{\delta_{\rm scr}}
\left(
\begin{aligned}
G^{(\rm reg)}_{\varepsilon\kappa}
\\
F^{(\rm reg)}_{\varepsilon\kappa}
\end{aligned}
\right)
+
\sin{\delta_{\rm scr}}
\left(
\begin{aligned}
G^{(\rm irr)}_{\varepsilon\kappa}
\\
F^{(\rm irr)}_{\varepsilon\kappa}
\end{aligned}
\right)
\end{equation}
where the phase shift $\delta_{\mathrm{scr}}$ is induced by the short-range screening potential.
The regular and irregular at origin solutions can be expressed in terms of the Coulomb functions~\cite{Abramovitz, Salvat_1995}
\begin{equation}
\left(
\begin{aligned}
G^{(\rm reg)}_{\varepsilon\kappa}
\\
F^{(\rm reg)}_{\varepsilon\kappa}
\end{aligned}
\right)
=
N
\left(
\begin{aligned}
p
\sqrt{\gamma^2 + \nu^2}
(\gamma+\kappa)
F_\gamma(-\nu, pr)
+
\alpha Z_{\rm as}
(\varepsilon\kappa - \gamma m_e)
F_{\gamma-1}(-\nu, pr)
\\
\alpha Z_{\rm as}
p
\sqrt{\gamma^2 + \nu^2}
F_{\gamma}(-\nu, pr)
+
(\gamma + \kappa)
(\varepsilon\kappa - \gamma m_e)
F_{\gamma -1}(-\nu, pr)
\end{aligned}
\right),
\label{eq:reg_solution}
\end{equation}
\begin{equation}
\left(
\begin{aligned}
G^{(\rm irr)}_{\varepsilon\kappa}
\\
F^{(\rm irr)}_{\varepsilon\kappa}
\end{aligned}
\right)
=
N
\left(
\begin{aligned}
p
\sqrt{\gamma^2 + \nu^2}
(\gamma+\kappa)
G_\gamma(-\nu, pr)
+
\alpha Z_{\rm as}
(\varepsilon\kappa - \gamma m_e)
G_{\gamma-1}(-\nu, pr)
\\
\alpha Z_{\rm as}
p
\sqrt{\gamma^2 + \nu^2}
G_{\gamma}(-\nu, pr)
+
(\gamma + \kappa)
(\varepsilon\kappa - \gamma m_e)
G_{\gamma -1}(-\nu, pr)
\end{aligned}
\right).
\label{eq:irreg_solution}
\end{equation}
Here $N$ is the normalization factor defined as
\begin{equation}
N
=
\sqrt{\frac{\varepsilon + 1}{\pi p}}
\frac{1}{\gamma\sqrt{p^2
(\gamma+\kappa)^2
+
\alpha^2 Z_{\rm as}^2
\left(
\varepsilon+1
\right)^2}}
\label{eq:normalization_factor}
\end{equation}
with $\gamma = \sqrt{\kappa^2 - (\alpha Z_{\rm as})^2}$ and
the Sommerfeld parameter $\nu = \alpha Z_{\rm as}\varepsilon/p$.
The Coulomb functions in~(\ref{eq:reg_solution}) and~(\ref{eq:irreg_solution}) can be expressed through the Whittaker functions of the second kind $W_{\alpha,\beta}$ as follows~\cite{Abramovitz}
\begin{equation}
F_\lambda(\eta, \rho)
=
\frac{H_\lambda^+(\eta, \rho) - H_\lambda^-(\eta, \rho)}{2i},
\quad
G_\lambda(\eta, \rho)
=
\frac{H_\lambda^+(\eta, \rho) + H_\lambda^-(\eta, \rho)}{2},
\label{eq:coul_functions}
\end{equation}
\begin{equation}
H^\pm_\lambda(\eta, \rho)
=
e^{\pi\eta/2}
e^{\mp i\left[ \frac{\pi\lambda}{2}-\sigma_\lambda(\eta) \right]}
W_{\mp i\eta,\,\lambda+1/2}(\mp2i\rho)
\label{eq:whittaker_functions}
\end{equation}
with $\sigma_\lambda(\eta)$ standing for the argument of the Euler's Gamma function $\Gamma(\lambda + 1 + i\eta)$~\cite{Abramovitz}.
\\
\indent
Since the integrand in~(\ref{eq:radial_integral}) is an oscillating function, we apply the rotation of the integration path into the complex plane.
For this purpose, the integral $P_{JL}(i,f)$ is divided into two parts:
\begin{equation}
P_{JL}(i,f)
=
P_{JL}^{(R)}(i,f)
+
P_{JL}^{(C)}(i,f),
\end{equation}
where the first part
\begin{equation}
P_{JL}^{(R)}(i,f)
=
\int\limits_0^{R}dr
j_L(\omega r)
\Big[
S_{JL}(\kappa_i, -\kappa_f)
G_{\varepsilon_i\kappa_i}(r)
F_{\varepsilon_f\kappa_f}(r)
-
S_{JL}(-\kappa_i, \kappa_f)
G_{\varepsilon_f\kappa_f}(r)
F_{\varepsilon_i\kappa_i}(r)
\Big]
\label{eq:radial_integral_R}
\end{equation}
is calculated numerically utilizing functions and subroutines from the modified \texttt{RADIAL} package~\cite{Salvat_1995}.
The remaining part
\begin{equation}
P_{JL}^{(C)}(i,f)
=
\int\limits_R^{+\infty}dr
j_L(\omega r)
\Big[
S_{JL}(\kappa_i, -\kappa_f)
G_{\varepsilon_i\kappa_i}(r)
F_{\varepsilon_f\kappa_f}(r)
-
S_{JL}(-\kappa_i, \kappa_f)
G_{\varepsilon_f\kappa_f}(r)
F_{\varepsilon_i\kappa_i}(r)
\Big]
\label{eq:radial_integral_C}
\end{equation}
is a linear combination of integrals
\begin{equation}
I
\equiv
I_{\alpha_i,\beta_i;\;\alpha_f,\beta_f;\; L}(s_i, s_f)
=
\int\limits_R^{+\infty}dr
j_L(\omega r)W_{\alpha_i,\beta_i}(2ip_is_ir)W_{\alpha_f,\beta_f}(2ip_fs_fr),
\label{eq:small_radial_integral_C}
\end{equation}
where $s_i$ and $s_f$ equal to $+1$ or $-1$.
The point $R$ is chosen so that for $r>R$ the asymptotics of all appearing Whittaker functions
\begin{equation}
W_{\alpha, \beta}(z)
\xrightarrow[|z|\rightarrow +\infty]{}
e^{-z/2}
z^{\alpha}
\left[
\sum\limits_{n=0}^{N-1}\frac{(1/2-\alpha+\beta)_n(1/2-\alpha-\beta)_n}{n!}(-z)^{-n}
+
O(|z|^{-N})
\right]
\label{eq:whittaker_asymp}
\end{equation}
converge with the required accuracy.
Here $(x)_n = x(x+1)\ldots(x+n-1)$ stands for the Pochhammer's symbol~\cite{Abramovitz}.
It can be seen from~(\ref{eq:whittaker_asymp}) that the Whittaker functions $W_{\alpha, \beta}(2ipr)$ and $W_{\alpha, \beta}(-2ipr)$ are regular in the lower and upper halfs of the complex $r$ plane, respectively.
They both decrease exponentially with $r$ moving away from the real axis.
Substituting~(\ref{eq:whittaker_asymp}) into~(\ref{eq:small_radial_integral_C}) and utilizing the explicit form of Bessel functions~\cite{Abramovitz}
\begin{equation}
j_L(z)
=
\sum\limits_{s=\pm 1}
\sum\limits_{n=0}^{L}
(-is)^{L+1-n}
\frac{(L+n)!}{n!(L-n)!}
\frac{e^{isz}}{(2z)^{n+1}},
\end{equation}
we arrive at
\begin{equation}
I
=
\int\limits_R^{+\infty}dr
r^{\alpha_i + \alpha_f -1}
\sum\limits_{s=\pm 1}
e^{i(s\omega-s_ip_i-s_fp_f)r}
\sum\limits_{n\geq 0}
\frac{c_n}{r^n}.
\label{eq:small_radial_integral_C_exp}
\end{equation}
Since for the bremsstrahlung process $p_i > p_f + \omega$, we perform the $\pi/2$-rotation of the integration path to the upper half of the complex $r$ plane for $s_i = -1$ and to the lower one for $s_i = +1$.
Employing the change of variables $r = R+is_i\xi$ gives us the converging factor
$e^{-p_i\xi}$ in each integral $I$.
%
%


\begin{thebibliography}{99}
%
%
\bibitem{Mangiarotti_RPC141_312:2017}
A.~Mangiarotti and M.~N.~Martins,
Rad. Phys. Chem. {\bf 141}, 312 (2017).
%
\bibitem{Jakubassa_arXiv_2103.06034}
D.~H.~Jakubassa-Amundsen,
arXiv:2103.06034.
%
%
\bibitem{TsengPratt_PRA3_100:1971}
H.~K.~Tseng and R.~H.~Pratt,
Phys. Rev. A {\bf 3}, 100 (1971).
%
\bibitem{TsengPratt_PRA7_1502:1973}
H.~K.~Tseng and R.~H.~Pratt,
Phys. Rev. A {\bf 7}, 1502 (1973).
%
%
\bibitem{Yerokhin_PRA82_062702:2010}
V.~A.~Yerokhin and A.~Surzhykov,
Phys. Rev. A {\bf 82}, 062702 (2010).
%
\bibitem{Tashenov_PRA87_022707:2013}
S.~Tashenov {\it et al.},
Phys. Rev. A {\bf 87}, 022707 (2013).
%
\bibitem{Muller_PRA90_032707:2014}
R.~A.~M\"uller, V.~A.~Yerokhin, and A.~Surzhykov,
Phys. Rev. A {\bf 90}, 032707 (2014).
%
\bibitem{Poskus_CPC04_030:2018}
Andrius~Po{\v s}kus,
Comput. Phys. Comm. {\bf 232}, 237 (2018).
%
\bibitem{Poskus_ADN129_101277:2019}
Andrius~Po{\v s}kus,
At. Data Nucl. Data Tables {\bf 129-130}, 101277 (2019).
%
\bibitem{Jakubassa_PRA100_032703:2019}
D.~H.~Jakubassa-Amundsen and A.~Mangiarotti,
Phys. Rev. A {\bf 100}, 032703 (2019).
%
\bibitem{Jakubassa_PRA82_042714:2010}
D.~H.~Jakubassa-Amundsen,
Phys. Rev. A {\bf 82}, 042714 (2010).
%
\bibitem{Jakubassa_PRA98_062715:2018}
D.~H.~Jakubassa-Amundsen,
Phys. Rev. A {\bf 98}, 062715 (2018).
%
\bibitem{Jakubassa_JPG47_075102:2020}
D.~H.~Jakubassa-Amundsen,
J. Phys. G: Nucl. Part. Phys. {\bf 47}, 075102 (2020).
%
%
\bibitem{Jakubassa_PRA93_052716:2016}
D.~H.~Jakubassa-Amundsen,
Phys. Rev. A {\bf 93}, 052716 (2016).
%
\bibitem{Nofal_PRL99_163201:2007}
M.~Nofal {\it et al.},
Phys. Rev. Lett {\bf 99}, 163201 (2007).
%
\bibitem{Hillenbrand_PRA90_022707:2014}
P.-M.~Hillenbrand {\it et al.},
Phys. Rev. A {\bf 90}, 022707 (2014).
%
\bibitem{Hillenbrand_PRA101_022708:2020}
P.-M.~Hillenbrand {\it et al.},
Phys. Rev. A {\bf 101}, 022708 (2020).
%
\bibitem{Rose_1961}
M.~E.~Rose,
{\it Relativistic Electron Theory}, (Wiley, New York, 1961).
%
\bibitem{Pratt_RMP45_273:1973}
R.~H.~Pratt, A.~Ron, and H.~K.~Tseng,
Rev. Mod. Phys. {\bf 45}, 273 (1973); {\bf 45}, 663(E) (1973).
%
\bibitem{Eichler_1995}
J.~Eichler and W.~Meyerhof,
{\it Relativistic Atomic Collisions} (Academic, San Diego, 1995).
%
\bibitem{Rose_1957}
M.~E.~Rose,
{\it Elementary Theory of Angular Momentum}, (Wiley, New York, 1957).
%
\bibitem{Varshalovich}
D.~A.~Varshalovich, A.~N.~Moskalev, and V.~K.~Khersonskii,
{\it  Quantum Theory of Angular Momentum},
(World Scientific, Singapore, 1988).
%
\bibitem{Abramovitz}
{\it Handbook of Mathematical Functions},
edited by M.~Abramovitz and I.~A.~Stegun
(U. S. Govt. Printing Office, Washington, D.C., 1964).
%
\bibitem{PerdewZunger}
J.~P.~Perdew and A.~Zunger,
Phys. Rev. B {\bf 23}, 5048 (1981).
%
\bibitem{Salvat_1987}
F.~Salvat, J.D.~Martinez, R.~Mayol, and J.~Parellada,
Phys. Rev. A {\bf 36}, 467 (1987).
%
\bibitem{Salvat_1991}
F.~Salvat,
Phys. Rev. A {\bf 43}, 578 (1991).
%
\bibitem{Jakubassa_JPB36_1971:2003}
D.~H.~Jakubassa-Amundsen,
J. Phys. B {\bf 36}, 1971 (2003).
%
\bibitem{Jakubassa_RPC75_1319:2006}
D.~H.~Jakubassa-Amundsen,
Radiat. Phys. Chem. {\bf 75}, 1319 (2006).
%
\bibitem{Jakubassa_JPB40_2719:2007}
D.~H.~Jakubassa-Amundsen,
J.Phys. B {\bf 40}, 2719 (2007).
%
\bibitem{Jakubassa_EPJD41_267:2007}
D.~H.~Jakubassa-Amundsen,
Eur. Phys. J. D {\bf 41}, 267 (2007).
%
\bibitem{Eichler_PR439_1:2007}
J.~Eichler and T.~St\"ohlker, 
Phys. Rep. {\bf 439}, 1 (2007).
%
\bibitem{Salvat_1995}
F.~Salvat, J.~M.~Fern\'andez-Varea, and W.~Williamson~Jr.,
Comput. Phys. Commun. {\bf 90}, 151 (1995).
%
%
%
\end{thebibliography}
\end{document}